\documentclass[fleqn,10pt]{wlscirep}

\title{Challenges and constraints of dynamically emerged source and sink in atomtronic circuits: From closed-system to open-system approaches}

\author[1]{Chen-Yen Lai}
\author[1]{Chih-Chun Chien}
\affil[1]{University of California Merced, School of Natural Sciences, Merced, 95343, USA.}

\begin{abstract}
While batteries offer electronic source and sink in electronic devices, atomic analogues of source and sink and their theoretical descriptions have been a challenge in cold-atom systems.
Here we consider dynamically emerged local potentials as controllable source and sink for bosonic atoms.
Although a sink potential can collect bosons in equlibrium and indicate its usefulness in the adiabatic limit, sudden switching of the potential exhibits low effectiveness in pushing bosons into it.
This is due to conservation of energy and particle in isolated systems such as cold atoms.
By varying the potential depth and interaction strength, the systems can further exhibit averse response, where a deeper emerged potential attracts less bosonic atoms into it.
To explore possibilities for improving the effectiveness, we investigate what types of system-environment coupling can help bring bosons into a dynamically emerged sink, and a Lindblad operator corresponding to local cooling is found to serve the purpose.
\end{abstract}

\begin{document}
\flushbottom
\maketitle

\section*{Introduction}
Recent advances in trapping and manipulating ultracold atoms in magnetic or optical potentials have brought new tools for studying non-equilibrium phenomena of many-body systems via quantum simulations~\cite{Cirac:2012jj,Georgescu:2014bg}.
In contrast to conventional solid state materials, ultracold atoms provide more flexibility in their structures and are controllable over a broad range of parameters such as interactions and temperature ~\cite{Chin:2010kl,Pethick:2010gy}.
Those new techniques also provide opportunities for testing and verifying theories of transport properties in solid state devices and cold atom systems~\cite{Labouvie:2015dx,Bushong:2005kx,Chien:2014uf,Peotta:2014woa,Chien:2012cy,Chien:2013en,Chien:2015kc,GallegoMarcos:2014fb}.
Recently, the concept of atomtronics~\cite{Amico:2015tk,Olsen:2015hd} has drawn intense  attention due to intriguing experimental and theoretical studies, including atomic SQUID~\cite{Wang:2015il,Arwas:2016ce,Safaei:2016uw,Mathey:2016iv,Astafiev:2012bi}, transistor~\cite{Caliga:2016fq}, capacitor~\cite{Li:2016ux}, and open quantum systems~\cite{Pepino:2010jx,Barontini:2013ub,Daley:2014hc}.
There is a bright future for atomtronics, and here we will address a challenging issue on  driving atoms in atomtronic circuits via local manipulations.

While particle reservoirs like batteries play the role of source or sink in conventional electronic systems, atomic analogues of particle source or sink for atomtronics are highly desired. However, due to charge neutrality of atoms, one needs creative ways for supplying or removing atoms.
It is possible to use atoms from a nearby trap as a source~\cite{Zozulya:2013cp} and remove atoms using photon or electron beams, which acts as a sink~\cite{Caliga:2012ul,Labouvie:2015dx,Barontini:2013ub}.
Nevertheless, programmable atomtronic circuits may need dynamically generated sources or sinks.
We will investigate whether local manipulations of the potential in a small region can act as a source or sink effectively in isolated systems modeling cold-atoms.
Such a scheme is more suitable for bosons since Pauli exclusion principle may limit the amount of fermions allowed in a narrow region.
For noninteracting and weakly interacting bosons, the ground state corresponds to a congregate of bosons in the deep potential.
However, we will show that the quantum nature and energy conservation severely compromise the effectiveness of a dynamic sink from a sudden emergence of a deep potential.
The Bose-Hubbard model (BHM) will be implemented and we alter the onsite energy of a selected single site to generate a sink.
In equilibrium, a mean-field estimation of the maximal amount of bosons attracted into the sink qualitatively agrees with numerical simulations.
The simulations show promising results in equilibrium with a large fraction of atoms in the sink.
This ensures the effectiveness of local deep potentials as atomic sinks in the adiabatic limit where the potential changes slowly.

Atomtronic circuits, nevertheless, are expected to operate within finite durations.
The finite hold time of atomic clouds also restricts the switching time of atomtronic devices.
Therefore, we explore the opposite limit where a local potential is suddenly changed and analyze its effectiveness as a dynamical sink.
As one will see shortly, a dynamically emerged source or sink acts poorly in providing or collecting quantum particles.
The origin of the ineffectiveness comes from the wave nature of quantum particles and energy conservation.
By ramping the sink potential to deeper depth, the number of bosons attracted into the sink may even decrease, and the system exhibits averse response reminiscent of the negative differential conductivity in electronic and atomic systems~\cite{Conwell:2008kf,Labouvie:2015dx}, where a stronger driving field leads to a smaller current.
The same conclusions are reached in a continuum model summarized in the Supplementary Information.

In order to explore ways for improving the effectiveness of dynamic source and sink for cold atoms, we relax the isolated-system condition by consider environmental effects and extend the theoretical description to an open-system approach.
For atomtronic systems, external perturbations by light or atoms need to be introduced for significant environmental effects.
For an open quantum system, one may describe the dynamics using the quantum master equation approach~\cite{Pepino:2010jx}, which has been extremely successful in quantum optics~\cite{Gardiner:2000kq,breuer2007theory,weiss2012quantum}.
We followed this method and analyzed the effects of different Lindblad operators on bringing the particles into a dynamically generated sink.
While a popular form of Lindblad operator in studying decoherence~\cite{Daley:2014hc,Pichler:2010cs,Poletti:2012di,Schachenmayer:2014fm} does not improve the amount of particles drawn into the sink, we explore a particular Lindblad operator inspired by a study of Bose-Einstein condensate formation~\cite{Griessner:2006hg,Griessner:2007hf,Daley:2004be} and find such an intervention draws substantially more particles into the sink.
Implications of this particular Lindblad operator and possible experimental connections will be discussed.
The quantum master equation approach complements the shortcut-to-adiabaticity approach~\cite{Torrontegui:2013ij}, where additional time-dependent deformations of potentials bring the system to its adiabatic limit.

\section*{Results}

\subsection*{Isolated quantum system}
First, we study Bose gases in a one dimensional lattice potential with tunable onsite energy of selected sites.
The system may be described by a single-band BHM model, whose Hamiltonian is given by
\begin{equation}\label{eq:bhm}
	\mathcal{H}=-J\sum_{\langle i,j\rangle}b^\dagger_ib_j-\sum_iV_i(t)n_i +\frac{U}{2}\sum_in_i(n_i-1).
\end{equation}
Here $b^\dagger_i$ $(b_i)$ is the boson creation (annihilation) operator on lattice site $i$, $n_i\!=\! b^\dagger_ib_i$ is the boson number operator on site $i$, and $\langle i,j\rangle$ represents nearest neighbors.
We set $\hbar\!=\! 1$ and the time unit is $t_0\!=\! \hbar/J$.
To simulate different setups with a sink, a source, or a combination of both, we consider different sequences of the time-dependent local potential energy, $V_i(t)$.

The ground state with or without a sink or a source can be found by the exact diagonalization (ED) method, and the dynamics can be monitored by using a similar technique.
We simulate small systems up to $L=13$ lattice sites and $N=11$ bosons.
For larger systems, the ED method is less practical and we rely on the density matrix renormalization group~\cite{White:1993fb,White:1992ie,Schollwock:2011gl} (DMRG) method.

\subsubsection*{Equilibrium ground state with a sink potential}

\begin{figure}[hb]
	\centering
	\includegraphics[width=0.8\linewidth]{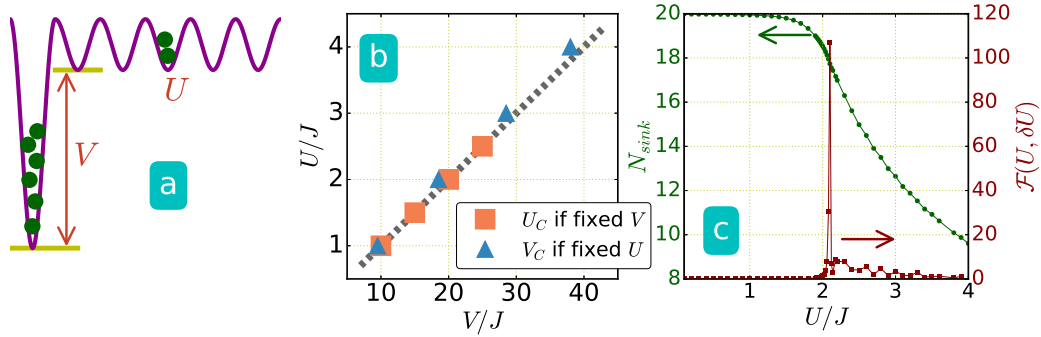}
	\caption{
	(a) Illustration of a single sink potential located at one edge of a lattice system in equilibrium.
	(b) ED results of systems with $L\!=\!13$ sites and $N\!=\!11$ bosons.
	The critical values of $U_c$ and $V_c$ are determined from the fidelity metric by fixing $V$ (square) and $U$ (triangle), respectively.
	The dashed line shows the critical values from the mean field approximation, which agrees quantitatively with the ED results (symbols).
	(c) Fidelity metric and number of particles inside the sink for a system with $L\!=\!61$ sites, particle number $N\!=\!20$, and potential depth $V\!=\!38J$ simulated by the DMRG method.
	}
	\label{fig:eqm}
\end{figure}

We first investigate the ground state in the presence of a sink modeled by a deep potential on one site, as illustrated in Fig.~\ref{fig:eqm}a.
For the noninteracting case with $U=0$, it can be shown that a bound state exists when a single site potential is deeper than the half bandwidth $4J$.~\cite{Anderson:1958fz}.
For a system with $L\!=\!13$ and $N\!=\!11$, we use the ED to simulate the ground state wavefunction with a sink at site $i$ ($V_{i}\! =\! V$) and $V_{j\neq i}=0$.
In order to deal with finite size effects, we calculate the fidelity metric defined as
\begin{equation}
	\mathcal{F}(V, \delta V) = \frac{2}{L}\frac{1-F(V, \delta V)}{(\delta V)^2},
\end{equation}
where the fidelity, $F(V, \delta V)\!=\!\langle\Psi_0(V)|\Psi_0(V+\delta V)\rangle$, is the overlap between two normalized ground states obtained with a small change in the parameter.
For finite-size systems, a decrease in the fidelity is a precursor to a crossover or quantum phase transition~\cite{CamposVenuti:2008gl,Varney:2010ej}, and that corresponds to a peak in the fidelity metric.
In our simulations we vary the potential depth when evaluating the fidelity.
For a fixed potential depth $V$ (or coupling constant $U$), we vary the interaction (or potential depth) and determine the critical point between the ground state with all particles in the sink and the ground state with one particle outside the sink.
The fidelity metric shows a sharp peak when the configuration with all particles in the sink is no longer stable.

The critical interaction strength (or potential depth) when the configuration with all particles in the sink is no longer stable is shown in Fig.~\ref{fig:eqm}b, which agrees well with a mean-field analysis shown in the  Supplementary Information.
A continuum model is also analyzed in the Supplementary Information and the results converge to the same conclusions.
We also use the DMRG to study a larger system with $L\! =\! 61$ sites, $N\! =\! 20$ bosons, and the sink potential $V\!=\!38J$.
The fidelity metric shows a peak in Fig.~\ref{fig:eqm}c around $U\!\approx\! 2.1J$, which is closed to the mean-field prediction of $U_c\!=\! 2J$ when the ground state with all particles in the sink is no longer stable.
In general, the number of particles in the sink potential decreases as the interaction becomes stronger, which implies that BHM may be driven across the Mott insulator-superfluid transition by manipulating the potential on one single site with suitable filling~\cite{Deng:2014th}.

\subsubsection*{System evolution with dynamically emerged sink}
\begin{figure}[ht]
	\centering
	\includegraphics[width=0.9\linewidth]{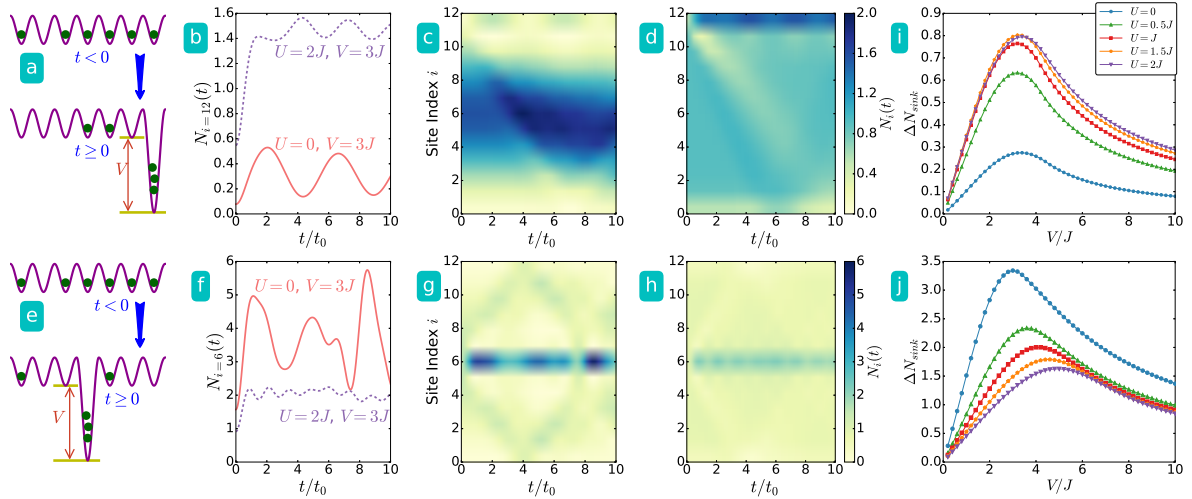}
	\caption{
		Sudden emergence of a sink at (a) the left edge and (e) the center of the system with size $L=13$ and $N=11$ bosons.
		The top and bottom rows show the results correspond to the setups of (a) and (e), respectively.
		(b) and (f): Time evolution of particle number in the sink potential.
		(c), (d) and (g), (h): Overall density evolution with $V\!=\!3J$ and $U\!=\!0$ ((c) and (g)) and with $V\!=\!3J$ and $U\!=\!2J$ ((d) and (h)).
		The upper (lower) color bars are for (c) and (d) ((g)-(h)).
		(i) Short time behavior of particle number in the dynamical sink potential versus potential depth under different interaction strength.
		(j) The same plot as (i) with sink located at center of the system.
	}
	\label{fig:sink}
\end{figure}

The equilibrium results suggest that a deep potential on one site may serve as a particle sink to collect bosons in the weakly interaction regime.
As the sink potential can be tuned very slowly, it is expected the system remains in the ground state after the sink potential emerges and all the particles stay in the sink if the ratio between the interaction and trap depth, $U\!/\!V$, falls below the critical value.
However, the adiabatic limit would be less relevant for designing scalable atomtronic devices.
The time scale of the emerged sink to satisfy the adiabatic limit in the noninteracting case can be shown to increase with the system size (see the Supplementary Information).
In the following we consider instantaneous switching of the sink or source potentials, including
1) a suddenly generated well potential at one edge or the center of the system as shown in Figs.~\ref{fig:sink}a and \ref{fig:sink}e,
2) an initial well potential at one edge is suddenly lifted while a well potential appears at the other edge as a source-sink combination shown in Fig.~\ref{fig:dss}a.
The interactions are assumed to be uniform.

The dynamics after the sink or source potential is turned on may be interpreted as a response theory, where the driving field corresponds to the sink or source potential and the response may be the particle number difference in the sink site.
Such a response theory is similar to the case where a magnetic field drives the magnetization or a mass current responds to a chemical potential difference.
We monitor the density distribution in real time with different values of the uniform coupling constant $U$ in Eq.~(\ref{eq:bhm}), and the results from the ED are shown in Fig.~\ref{fig:sink} for $L\!=\!13$ and $N\!=\!11$.
By quenching the sink potential to a constant value $V\!=\!3J$, we observe different dynamics with different interacting strength, but in general the dynamically generated sink attracts much less particles when compared to its equilibrium counterpart.
For example, in the case with a sink at the edge shown in Figs.~\ref{fig:sink}a-~\ref{fig:sink}d, less than $2\%$ of the total particles flow into the suddenly emerged sink.
The case with a sink at the center shown in Figs.~\ref{fig:sink}e-~\ref{fig:sink}h can attract more particles due to the initial inhomogeneous density distribution, but it is still far less than the equilibrium counterpart.
The difference between the case with a sink at one edge and the case with one at the center is that stronger interaction strength pushes more particles towards the edge in the initial state.
The results with different sink potential depth and various interaction strength are summarized in Figs.~\ref{fig:sink}i and \ref{fig:sink}j.
Here $\Delta N_{sink}$ is defined as the particle number difference in the sink between the initial value and the first peak in its evolution. (See Fig.~\ref{fig:sink}b for example.) The continuum model analyzed in the Supplementary Information also exhibits similar ineffectiveness of a dynamically emerged sink and dependence of the sink location.

The reason for the low efficiency of the dynamically generated sink is mainly due to the conservation of energy in isolated systems such as cold atoms.
After the sink potential suddenly appears, the ground state of the initial uniform lattice becomes a relatively high-energy state of the new Hamiltonian with the sink potential.
The low-energy states in the presence of the sink should be those with particles localized inside the sink.
When one particle hops into the sink potential, it will lower the energy by an amount of the order of $V$.
Due to energy conservation, this energy loss has to be compensated by, for example, the kinetic or interaction energy.
For noninteracting gases, the kinetic energy per particle is constrained by the bandwidth $\mathcal{W}\!=\!4J$.
Therefore, it is impossible for particles to accumulate in the sink when the loss of potential energy is much larger than the bandwidth representing the kinetic energy.
Although similar analyses show that adding weakly repulsive interactions allows few more particles to flow into the sink potential, complexity arises in the strong interaction regime and will be discussed later.

For fixed interaction strength and relatively weak sink depth, the maximal amount of particles drawn into the sink increases as the sink potential is quenched to larger values, which indicates an improvement of the effectiveness of a dynamical sink.
However, the amount of particles in the sink decreases when the depth of the sink potential exceeds a critical value. This indicates that in the deep sink regime, the system exhibits averse response, where a deeper sink potential results in less particles in the sink.
This averse response is similar to the negative differential conductivity (NDC)~\cite{Conwell:2008kf}, where a stronger driving field leads to less current, and the NDC has been discussed and observed in cold-atom experiments~\cite{Chien:2013ef,Labouvie:2015dx}.

The issue on whether introducing interactions can improve the effectiveness of a dynamically emerged sink is complicated by several issues.
For example, if the quenched sink potential depth is fixed and the interactions are set to different values, the dynamics depends on the sink location because the initial density profiles change with interactions.
Moreover, increasing the interaction tend to reduce the maximal number of particles allowed in the sink due to the repulsion between particles.
In general, if the sink is located at one edge, stronger interactions can lead to more particles in the sink.
Fig.~\ref{fig:sink}i shows that there is an optimal potential depth for selected interaction strength.
For the case with a sink quenched at the center, the interactions do not provide observable improvement as shown in Fig.~\ref{fig:sink}j.
The help of effectiveness from the interactions disappears as the interaction energy exceeds a critical value when a site with multiple interacting bosons leads to huge interaction energy, and the number difference in the sink, $\Delta N_{sink}$, becomes insignificant.

\begin{figure}[ht]
	\centering
	\includegraphics[width=0.78\linewidth]{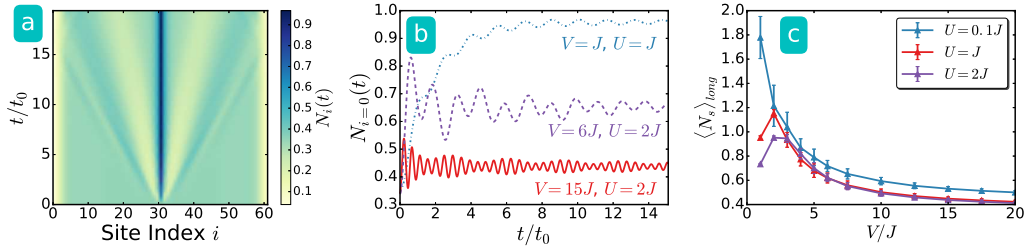}
	\caption{
	Sudden emergence of a sink at the center site $i\!=\!30$ for a lattice of $L\!=\!61$ sites and $N\!=\!20$ bosons calculated by the tDMRG method.
	(a) Evolution of the density profile for $U\!=\!J$ and $V\!=\!J$. Here we focus on the time before the particles bounce back from the edges.
	(b) Number of particles in the sink versus time for different interaction strength and potential depth.
	(c) Intermediate-time average ($10t_0\!-\!15t_0$) of the particle number in the sink.
	The error bar is due to statistical average.
	}
	\label{fig:dmrgsc}
\end{figure}

To simulate larger systems, we use the tDMRG with $L\!=\!61$ sites and $N\!=\!20$ bosons and study the intermediate-time behavior before the matter wave due to the sudden appearance of the sink potential bounces back from the edges and exhibits finite-size effects.
A light-cone structure can be observed in the time evolution of the density profile as shown in Fig.~\ref{fig:dmrgsc}a.
The effectiveness of the dynamically emerged sink, however, is not improved for larger systems due to conservation of energy.
The number of particles in the sink is shown in Fig.~\ref{fig:dmrgsc}b, which oscillates in time with a frequency depending on the quenched potential depth $V$.
The long-time behavior of this quantity tends to approach a stationary value, so we take its long-time average and plot it in Fig.~\ref{fig:dmrgsc}c.
Clearly, averse response showing a decreased number of particles in the sink as the sink potential increases is observable as the potential depth exceeds a critical value depending on the interaction.
Thus, the behavior of larger systems from the tDMRG qualitatively agrees with smaller systems calculated from the ED.

\subsubsection*{Transport in combined dynamic source and sink}
\begin{figure}[t]
	\centering
	\includegraphics[width=0.8\linewidth]{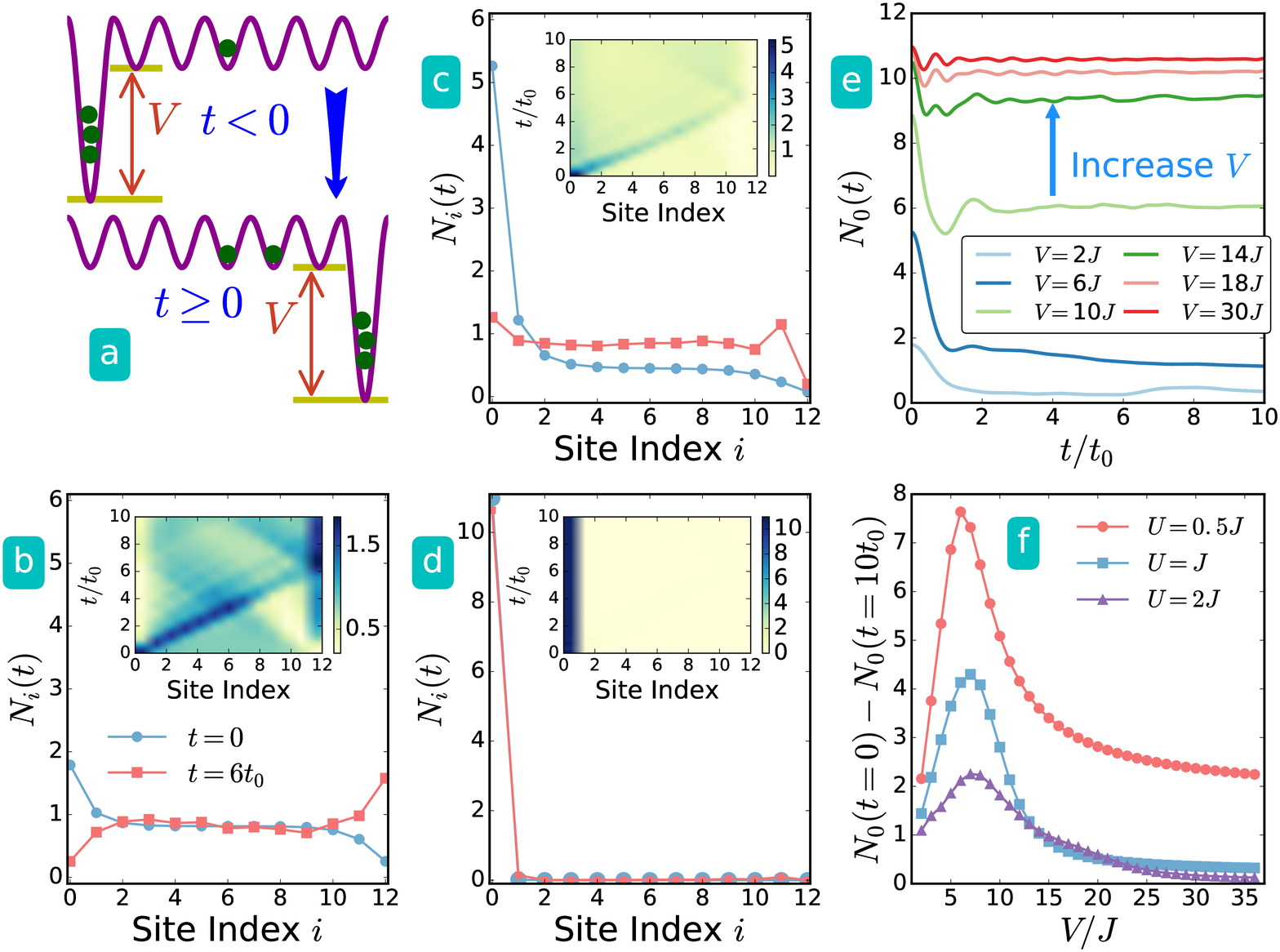}
	\caption{
		(a) Illustration of dynamically emerged source-sink combination.
		The potential well for the source is on the left most site while the sink appears on the right most site.
		In (b)-(f) we show the results with system size $L\!=\!13$ and $N\!=\!12$ particles.
		(b)-(d) Density distribution for two selected times, $t\!=\!0$ (blue circle) and $t\!=\!6t_0$ (red square).
		The inset shows the density contour versus time with interaction strength $U\!=\!J$ and potential depth (b) $V\!=\!2J$, (c) $V\!=\!6J$, and (d) $V\!=\!36J$.
		(e) Number of bosons on the source site versus time with $U\!=\!J$.
		(f) Particle number difference at the source site between the initial and later ($t\!=\!10t_0$) times versus $V$.
		Averse response manifests itself in (e) and (f) as larger depth drives less particles across the system.
	}
	\label{fig:dss}
\end{figure}

Next, we consider a system with a potential well initially at one end.
Then, another sink potential appears on the opposite end and the initial potential vanished at the same time as illustrated in Fig.~\ref{fig:dss}a.
The initial potential well may be interpreted as a particle source because the ground state has an initial surplus of particles in the well which are pushed out and generates a mass current.
This setup may be interpreted as a pair of dynamically generated source and sink, and the results are summarized in Fig.~\ref{fig:dss}b-\ref{fig:dss}f.
Initially, the well for reserving particles is located at the left end (the source site) with depth $V$, then the potential suddenly rises to zero while the sink potential appears on the right end (the sink site) with the same depth.
We begin with the case where the interaction $U$ is fixed and the depth of the source and sink potentials is varied to check if this dynamical process can induce a current through the system.
This is indeed the case as one can see in Fig.~\ref{fig:dss}b and~\ref{fig:dss}c, where the initial surplus of particles on the left is transferred to the right at a later time.
An interesting finding in the combined source and sink setup is illustrated in Fig.~\ref{fig:dss}b.
One can see that a few particles are transferred from the left source site to the right sink site.
The number of particle transported is sensitive to the ratio between the sink potential depth and the interaction strength.
Furthermore, the density evolution in the inset shows that the particles stay in the sink site after they arrive there.
Thus, the combined dynamic source and sink shed light on controlling few-particle transport across a quantum device with strongly interacting particles, which may be more difficult to demonstrate in conventional solid state systems~\cite{Li:2014iq}.

As shown in Fig.~\ref{fig:dss}d with large potential depth, there is very little flow of particles from the source site (on the left edge) after the potential energy is lifted.
Therefore, observable few-particle transport only occurs when the initial potential depth is not too deep and there are only slightly more particles in the source site than in other sites.
When the depth exceeds a critical value depending on the interaction strength, fewer particles can flow even when more particles are initially in the source site, and this demonstrates another example of averse response.
Fig.~\ref{fig:dss}e shows the particle number at the source site as a function of time with different potential depth, and one can see the averse response when a deeper initial potential pushes less particles out of it after the potential is lifted.
This averse response is general for various interaction strength, as one can observe it in the difference of the initial and later-time particle numbers on the source site in Fig.~\ref{fig:dss}f.

By a similar analysis using energy conservation, the initial interaction energy of the particles on the source site can be much larger than the kinetic energy limited by the bandwidth $\mathcal{W}\!\sim\!4J$.
Thus, the system cannot compensate for the loss of interaction energy when particles leave the source site.
Recent experiments~\cite{Ronzheimer:2013dr,Hackermuller:2010cm} showing slowing down of particle transport in interacting bosonic and fermionic systems may be partly related to this energy conservation constraint.
We remark that here we consider a single-band BHM, but in a more general model there may be more than one energy levels inside a deep potential.
Nevertheless, one may confine the analysis of a dynamic sink by focusing on the highest energy state in the sink potential (whose energy is still lower than the states outside the sink).
The lower-energy states in the sink are less relevant for satisfying energy conservation due to their larger energy differences with the initially uniform state.
Therefore, multi-state effects effectively reduce to the single-state potential considered here, and the results should be qualitatively the same.

\subsection*{Open Quantum System Approach}
So far the results show that it is very challenging to induce significant transport in an isolated system by only dynamically manipulating the potential in a small region.
It is possible, however, to introduce environmental effects via external light-atom or atom-atom interactions.
Here we will relax the isolation condition and investigate whether transport can be enhanced by external effects.
A system under external influence may be modeled by open-system approaches, which have been studied extensively and discussed in many areas of physics, especially quantum optics~\cite{scully1997quantum,breuer2007theory,weiss2012quantum} and spin systems~\cite{Pepino:2010jx,Daley:2014hc,Muller:2012ir,Barontini:2013ub}.
Here we implement a commonly used approach in open quantum systems and summarize the key approximations which simplify the description of dynamics in the Supplementary Information.

After considering the Born-Markov approximation, we arrive at the master equation, which takes the Lindblad form~\cite{Gardiner:2000kq,breuer2007theory,weiss2012quantum}
\begin{equation}
	\frac{d\rho_{\text{s}}}{dt}=-\frac{i}{\hbar}[\mathcal{H}, \rho_{\text{s}}]+\gamma\sum_j\left[L_j \rho_{\text{s}} L_j^\dagger-\frac{1}{2}\{\rho_{\text{s}},L_j^\dagger L_j\}\right].
\end{equation}
Here $L_j$ is a Lindblad operator, and $\gamma$ is a parameter characterizing the coupling between the system and environment.
We have tested some Lindblad operators discussed in the literature. For example, the local density operators have been implemented with the set of the Lindblad operators $\{L_j\}$ set to $\{n_l, l\!\in\![1,L]\}$.
This is one type of commonly used operators for introducing dissipation and decoherence, for example, in  dephasing of hard-core lattice bosons~\cite{Daley:2014hc,Labouvie:2015dx}.
This process leads to localization of atoms on each single site, which corresponds to spreading of the localized particles over all possible states in quasi-momentum space.
Hence, the kinetic energy decays to zero during the process~\cite{Daley:2014hc,Diehl:2008ha}.
According to a recent study~\cite{Olsen:2016tg}, using local density operators as the Lindblad operator can overcome the NDC of interacting bosons in a three-site potential.
In that study, the initial state corresponds to an inhomogeneous density distribution with a single empty site and two adjacent sites with finite density.
Since the local density operators favor a uniformly distributed density in the Lindblad equation, this process can dynamically fill the empty site as the system experiences decoherence.

\begin{figure}[ht]
	\centering
	\includegraphics[width=0.8\linewidth]{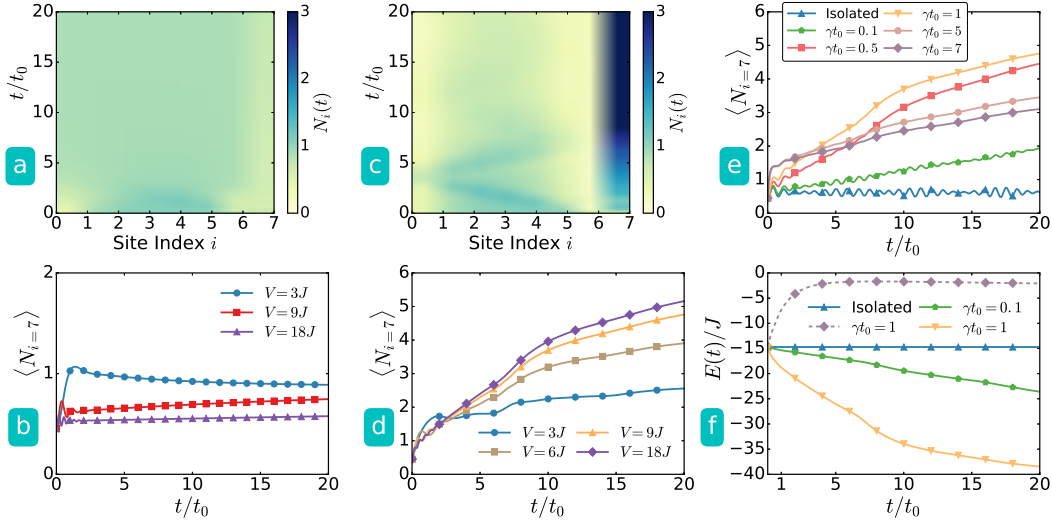}
	\caption{
	Open quantum system approach for a system with $L\!=\!8$ sites and $N\!=\!7$ bosons using the ED method.
	(a)-(b) are from Lindblad operators using the particle-number operators, $\{b^{\dagger}_{i}b_i, i=1,\cdots, L\}$, and (c)-(e) are from the local Lindblad operator $\{b^\dagger_kb_{k-1}, k\!=\!\text{sink site}\}$.
	(a) Time evolution of density profile of the case with sink depth $V\!=\!3J$.
	(b) Number of particles in the sink site ($i=7$) versus time with $U\!=\!J$. The system-environment coupling is set to $\gamma\!=\!1/t_0$ in (a) and (b).
	The particle number operators used as the Lindblad operators are thus not effective in attracting particles into a dynamic sink.
	(c) Time evolution of density profile of the case with $V\!=\!3J$.
	(d) Number of particles in the sink site versus time with $U\!=\!J$. $\gamma\!=\!1/t_0$ in (c) and (d).
	(e) Time evolution of the number of particles in the sink site for selected values of $\gamma$ with $U\!=\!J$ and $V\!=\!9J$.
	(f) Comparison of the system energy versus time for isolated system (triangular symbols), particle-number operators as the Lindblad operators (dashed line) and the local Lindblad operator with selected values of $\gamma$ (the lower two curves), where the sink potential is $V\!=\!9J$ and $U\!=\!J$.
	The local Lindblad operator is efficient in attracting particles into the dynamically generated sink.
	}
	\label{fig:LB}
\end{figure}

The local density operators, however, do not attract particles into a dynamically emerged sink because they favor uniform density distributions.
To verify whether a selected type of Lindblad operators can enhance the effectiveness of a dynamically emerged sink, we run simulations using the ED method for small systems.
To connect to the isolated-system results, we consider weak coupling between the system and environment in the sense that $\gamma$ is smaller than the rate $V/\hbar$ determined by the quenched sink potential.
The results according to the Lindblad master equation are summarized in Fig.~\ref{fig:LB}.

If the sink is quenched at one edge, the local density operators push some particles into the sink when they make the density uniform as shown in Fig.~\ref{fig:LB} a and b, but this type of Lindblad operators cannot drive the majority of bosons into the sink.
For the same reason, this process even reduces the particles in a dynamical sink quenched at the center of system because the initial ground state has higher density at the center and the particles spread out to reach a uniform density distribution.
Thus, the local density operators are useful for bringing an initially state with an inhomogeneous density distribution in real space into a final state with a uniform density distribution.
For a dynamically emerged sink, we are searching for Lindblad operators that work the other way around.

Another scheme~\cite{Griessner:2006hg,Griessner:2007hf} of Lindblad operators leads to a coherent driving of bosons into a Bose-Einstein condensate (BEC) by implementing the Lindblad operators $\{L_j\}$ with  $\{(b^\dagger_p+b^\dagger_q)(b_p-b_q), \text{where } p,q \text{ are neighboring sites}\}$.
This coherent driving is designed to produce a steady state~\cite{Diehl:2008ha} of BEC, which is the only dark state of the proposed Lindblad operators.
In other studies, this type of Lindblad operators is further extended to certain many-body systems to create pairing states~\cite{Kraus:2008jd,Daley:2014hc}.
Inspired by the latter scheme, we consider a Lindblad operator acting only on the sink site and its neighbor, so  $\{L_j\}$ is set to $\{b^\dagger_kb_{k-1}, k\!=\!\text{sink site}\}$.
Fig.~\ref{fig:LB}c clearly shows that this type of Lindblad operators not only helps the system draw more particles into the sink potential dynamically, but it also minimizes the back scattering due to the quenched potential.
As the potential depth increases, the sink accommodates more particles and the number of particles in the sink can be as large as $80\%$ of the total particles as shown in Fig.~\ref{fig:LB}d.

Therefore, the local Lindblad operator is efficient in bringing the system into the vicinity of its equilibrium configuration of the new Hamiltonian with an emerged sink.
Interestingly, the dependence of the amount of particles attracted into the sink in the long-time limit on the system-environment coupling is non-monotonic, as shown in Fig~\ref{fig:LB}e.
There is a maximal value of $\gamma$ where particles can be efficiently drawn into the sink. When the coupling increases further, the amount of particles attracted into the sink decreases.
Non-monotonic dependence of particle transport on system-environment coupling has also been discussed in fermionic systems allowing exchange of particles between the system and environment ~\cite{Velizhanin:2015hb}.

When more weakly-interacting particles are brought into a deep sink, the overall energy of the system should decrease due to the sink potential.
Indeed, as shown in Fig.~\ref{fig:LB}f the energy of the system decreases during the dynamics governed by the Lindblad equation with the local operator, which indicates a net flow of energy out of the system.
Thus, the local Lindblad operator may be considered as modeling a local cooling process, which may be of experimental interest as single-site cooling techniques have been developed recently~\cite{Grunzweig:2010dr,Kaufman:2012ft}.

Yet another route for improving the effectiveness of the dynamically generated sink is, at least theoretically, to approximate the system-environment interaction as a relaxation process in the Liouville-von Neumann equation~\cite{breuer2007theory}.
The equation of motion of the system is modeled as
\begin{equation}
	\frac{d\rho_s}{dt}=-\frac{i}{\hbar}[\mathcal{H}, \rho_s]-\frac{\rho_s-\rho_{eq}}{\tau_s}
\end{equation}
with $\rho_{eq}$ determined from the ground state of the final Hamiltonian with a dynamically emerged sink.
The relaxation time $\tau_s$ is usually treated as a phenomenological parameter.
The relaxation approximation, despite its simplicity, suffers some drawbacks.
For instance, the total particle number may not be strictly conserved during the dynamics.
In contrast, the two aforementioned Lindblad operators and their equations respect particle conservation during the evolution as shown in the Supplementary Information.
Secondly, the relaxation approximation may not guarantee the semi-positivity of the density matrix during the dynamics~\cite{breuer2007theory,weiss2012quantum}.
Moreover, this method requires a prior knowledge of the final ground state rather than a set of local Lindblad operators.
When considering atomtronic devices as assemblies of various local elements, the Lindblad master equation approach would be more versatile.

\section*{Discussion}
Energy and particle conservation in isolated systems like cold atoms impose constraints on their transport phenomena and lead to challenges on how to dynamically store or transfer particles.
Although we present results from one-dimensional systems, the mechanisms behind those phenomena should be general in higher dimensions.
In equilibrium, the systems show promising capability of accommodating particles in a sink potential.
The equilibrium results guarantee the functionality of a dynamically generated sink in the adiabatic limit, but it may not be particularly useful in atomtronic circuitry requiring short switching times.
For a sudden switch of the sink or source potential, our simulations show a lack of effectiveness to drive the particles into the equilibrium distribution, and this demonstrates another stark contrast between atomtronic and electronic systems~\cite{Chien:2015kc}.
A dynamically generated sink can even lead to averse response where an increase of the potential depth attracts less particles into it.
Nevertheless, few-particle transport could still be observable in interacting systems with  combined dynamical source and sink.

To explore how external effects can help improve the effectiveness of a dynamically emerged source or sink, we test the master-equation approach with various kinds of Lindblad operators.
While the commonly-used local density operators favor a uniform density distribution after time evolution, it is not helpful in the design of dynamical source and sink potentials.
Instead, a local Lindblad operator showing local cooling behavior is found to significantly improve the effectiveness of a dynamical sink.
This observation suggests that a combination of local cooling/heating and site-wise manipulations will have a bright future in making dynamical or programmable sinks or sources in atomtronic devices.

\section*{Method}

\subsection*{Exact diagonalization}
The L\'{a}czos procedure~\cite{Lanczos:1950wm,Lin:1993be} can calculate a few targeted states of a Hamiltonian which is sufficiently sparse, and a similar technique can be used to calculate real time dynamics.
The method uses the Krylov-space approach~\cite{Manmana:2005cg,Moler:2003fn} to approximate the time-evolution operator $\hat{\mathcal{U}}\!=\! e^{idt/\hbar\hat{\mathcal{H}}(t)}$, which evolves the wavefunction from time $t$ to time $t+dt$ according to
\begin{eqnarray}
	|\Psi(t+dt)\rangle&=&\hat{\mathcal{U}}|\Psi(t)\rangle\nonumber \\
	&\approx& \mathcal{V}_n(t)e^{-i\mathcal{T}_n(t)dt} \mathcal{V}^{T}_n(t)|\Psi(t)\rangle .
\end{eqnarray}
More specifically, the Krylov subspace spanned by the vectors
\begin{equation}
	\left\{|u_0\rangle, \mathcal{H}|u_0\rangle, \mathcal{H}^2|u_0\rangle, \cdots,\mathcal{H}^n|u_0\rangle\right\}
\end{equation}
are orthogonalized with respect to the previous two vectors in the set, which leads to the
L\'{a}nczos vectors
\begin{equation}
	|u_{j+1}\rangle=\mathcal{H}|u_j\rangle-\alpha_j|u_j\rangle-\beta_j^2|u_{j-1}\rangle
\end{equation}
with the coefficients $\alpha_j=\frac{\langle u_j|\mathcal{H}|u_j\rangle}{\langle u_j|u_j\rangle}$ and $\beta_j^2=\frac{\langle u_j|u_j\rangle}{\langle u_{j-1}|u_{j-1}\rangle}$.

For a given time $t$, we use $|\Psi(t)\rangle\!=\!|u_0\rangle$ as the first basis in the Krylov subspace.
The matrix $\mathcal{V}_n$ is composed of the L\'{a}nczos vectors in the form
\begin{equation}
	\mathcal{V}_n=\left(\begin{array}{cccc}
	\vdots & \vdots &  & \vdots \\
	|u_0\rangle & |u_1\rangle & \cdots & |u_{n-1}\rangle \\
	\vdots & \vdots &  & \vdots \\
	\end{array}\right).
\end{equation}
Thus, the Hamitonian can be expressed by the tridiagonal matrix
\begin{equation}
	\mathcal{T}_n=\left(\begin{array}{ccccc}
	\alpha_0 & \beta_1 & 0 & \cdots & \\
	\beta_1 & \alpha_1 & \beta_2 & 0 & \cdots \\
	0 & \beta_2 & \alpha_2 & \ddots & \\
	\vdots & 0 & \ddots & \ddots & \beta_n \\
	 &  &  & \beta_n & \alpha_n \\
	\end{array}\right).
\end{equation}
This procedure is exact if the number of the L\'{a}czos vector used, $n$, is equal to the dimension of the total Hilbert space of the Hamiltonian $\mathcal{H}$.
However, it is possible to obtain results with high accuracy by taking just a few L\'{a}czos vectors and a small $dt$, and the error of the Euclidean norm of the wavefunction is controllable~\cite{Manmana:2005cg,Hochbruck:2006cq}.
Here we use $20$ Lanczos vectors and a time step $dt\!=\!0.01t_0$, and the estimated error is around $10^{-10}$.

\subsection*{Density matrix renormalization group}
The tDMRG simulations have been applied to larger systems in and out of equilibrium.
For the simulations of equilibrium systems, we keep up to 150 states (bond dimension) and maintain the truncation error~\cite{Schollwock:2005jv} below $10^{-11}$.
For out-of-equilibrium dynamics, we decompose the evolution operator using the second-order Suzuki-Trotter formula~\cite{Vidal:2003ug,Vidal:2004jc} and evolve the ground state obtained from the static DMRG algorithm by the time-dependent DMRG~\cite{White:2004fd,Schollwock:2011gl,Lai:2008ez} (tDMRG).
During the simulations of time-dependent systems, the entanglement entropy increases drastically~\cite{Schollwock:2011gl}.
We manage to keep the truncation error below $10^{-8}$ but do not keep more than 1000 states.


\section*{Acknowledgments}
We are grateful to Malcolm Boshier, Roland Winston, Kevin Mitchell, Chen-Lung Hung, and David Weld for many useful discussions.
This work used the Extreme Science and Engineering Discovery Environment (XSEDE)~\cite{Towns:2014gd}, which is supported by National Science Foundation grant number ACI-1053575.

\section*{Author contributions statement}
C.C.C conceived the idea, and C.Y.L. performed the numerical calculation. C.C.C. and C.Y.L. wrote the manuscript and contributed equally to this work.

%

\clearpage
\appendix

\textbf{\Large Supplementary information for ``Challenges and constraints of dynamically emerged source and sink in atomtronic circuits: From closed-system to open-system approaches''}

\section{Mean Field Analysis of Equilibrium Sink}
When the coupling constant $U$ and sink-potential depth $V$ are large compared to the hopping coefficient $J$, we can estimate how many bosons are allowed in a sink potential by ignoring the kinetic energy.
For a system with $N$ particles, the on-site energy of all particles localized in the sink is approximated by
$E(N)\approx \frac{U}{2}N(N-1)-VN$.
We construct another state where $N_{out}\ll N$ particles are outside the sink, and its energy is approximated by
$E(N-N_{out})$.
The condition that the bosons are more stable to stay in the sink in this approximation is $E(N)-E(N-N_{out})<0$, which leads to
\begin{equation}
	\left(\frac{V}{U}\right)_c>\frac{2N-N_{out}-1}{2}.
\end{equation}

\section{Sink in continuum model in and out of equilibrium}
For the continuum model, the ground state and its dynamics may be studied by the Schr\"{o}dinger equation for noninteracting atoms or the mean-field Gross-Pitaevskii equation (GPE)~\cite{Gross:1961bx,Pitaevsk:fe} for weakly interacting bosons, as previously implemented in modeling coherent transport~\cite{Rab:2008kp,Bradly:2012fc}.
We will analyze a simplified model where a sink corresponds to a narrow square well inside a finite box, where a dilute quantum Bose gas in the weak-interaction regime can be described by a mean-field approach~\cite{Pethick:2010gy}.
At zero temperature, the condensate is described by an effective condensate wave function $\Phi(r,t)$.
The evolution of the condensate wave function in an external potential $V(r,t)$ is described by the GPE:
\begin{equation}
\left[-\frac{\hbar^2}{2m}\frac{d^2}{dx^2}+V_{ext}(x,t)+U_lN_b|\Phi|^2\right]\Phi = i\hbar\frac{\partial}{\partial t}\Phi ,
\end{equation}
where $m$ is the mass of the bosonic atom and $N_b$ is the number of bosons.
Here we solve the GPE with algorithms involving real- and imaginary-time propagation based on a split-step Crank-Nicolson method~\cite{Muruganandam:2009dq,Vudragovic:2012jz}, and follow Ref.~\cite{Cerimele:2000hs} to normalize the wavefunction with $\int dx |\Phi(x)|^2\!=\!1$.
The coupling constant $U_l\!=\! 4\pi\hbar^2a_s/m$ is determined by the two-body $s$-wave scattering length $a_s$.
The external potential $V_{ext}(x)$ corresponds to a narrow well and is set to simulate the equilibrium or dynamical sink.
A narrow, deep trap inside an overall harmonic trap has been realized in Ref.~\cite{Stellmer:2013wi}, and here we idealize the situation by considering square-well potentials.

\begin{figure}[t]
	\begin{center}
		\includegraphics[width=0.8\textwidth]{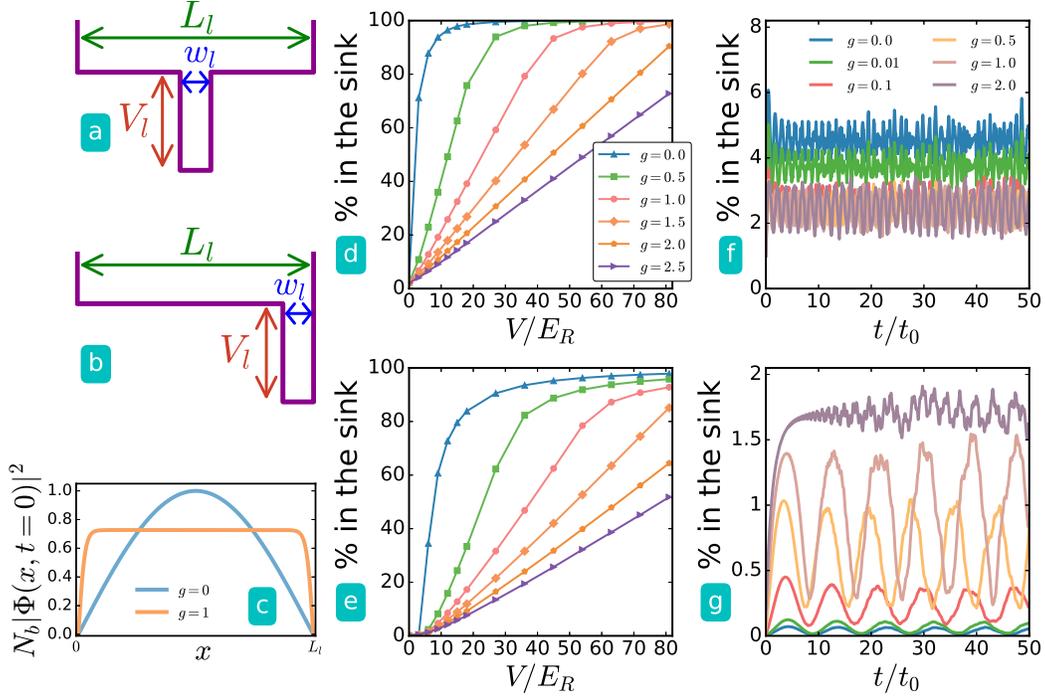}
		\caption{
			Illustration of the continuum model where a square-well sink is located at (a) the center and (b) the right edge of a box potential.
			(c) The initial density distributions with different interactions.
			(d) and (e): Fraction of bosons inside the sink in equilibrium as the sink depth varies.
			The results are from the GPE with $w_l\!/\!L_l\!=\!0.01$ and $N_b\!=\!25$ bosons for a sink located at (d) the center and (e) the right edge and different interaction strength $g$.
			(f) and (g): Fraction of bosons inside the sink as a function of time after a sink potential suddenly appears.
			The potential depth is $V_l\!/\!E_R\!=\!9$ and the sink is located at (f) the center and (g) the right edge.
			In (f), the data from $g\!=\!0.1$, $0.5$, $1.0$, and $2.0$ are all oscillating around $2.5\%$.
			The time unit here is $t_0=\hbar/E_R$.
			}
		\label{fig:GP}
	\end{center}
\end{figure}

The setups and their equilibrium results are shown in Fig.~\ref{fig:GP}a-\ref{fig:GP}e, where the system is confined in a one-dimensional box with length $L_l$, which is taken as the unit of length, and the particle number $N_b\!=\!50$.
We consider a square well potential of depth $V_l$ and width $w_l\ll L_l$ at the center or at one edge.
The reason we explore different locations of the sink is because the initial condensate wavefunction may not be uniform and the dynamics may be different. Moreover, the initial density varies with the interaction as illustrated in Fig.~\ref{fig:GP}c.
When presenting the results, however, we will focus on features that are not sensitive to the location of the sink.

We choose a narrow width $w_l\!=\!0.01L_l$ of the sink as shown in Fig.~\ref{fig:GP}.
For a non-interacting Bose gas at zero temperature, the number of bound states inside a square well is determined by the width and depth~\cite{Messiah:1999qm}.
For weakly interacting Bose gases with coupling constant $U_l\!=\!gE_R\Omega$ in equilibrium, less particles can be accommodated in the sink with larger $g$ due to the interaction energy, but the number of particles in the sink can be increased by increasing the depth of the sink potential.
Here $\Omega=L_l^3$ and $E_R\!=\!\pi^2\hbar^2/(2mL_l^2)$ is the recoil energy.

In the adiabatic limit~\cite{KatoTosio:1950jn,griffiths2005introduction} when the change of the sink potential is infinitely slow, the state remains in the ground state and the number of particles in the sink will eventually agrees with the equilibrium case.
However, the time required to approximate the adiabatic limit scales as $L_l^2$ and hinders the scalability of the device.
A similar constraint also applies to the lattice case.
In the following we will focus on setups with a sudden switch-on of a sink or source.
To simulate a dynamically emerged sink, the potential is uniform with $V_l(x,t<0)\!=\!0$ initially, then a quench to a deep sink potential leads to transport of atoms.

The suddenly emerged sink, however, does not work as expected when compared to its equilibrium counterpart.
Fig.~\ref{fig:GP}f shows the percentage of particles flowing into a dynamically emerged sink potential at the center, and Fig.~\ref{fig:GP}g shows the case for a sink at the right edge of the system.
Interestingly, in neither cases the maximal fraction of particles in the sink reaches $6\%$.
This low effectiveness of a dynamically emerged sink is a consequence of energy conservation.
The ground-state energy of the initial configuration without a sink is higher than that of the final configuration with a sink because without a sink the particles spread over the whole system while with a sink most particles tend to localize inside the sink to take advantage of the low potential energy.
In an isolated system such as cold atoms, there is no external dissipation to relax the system from the ground state of the initial Hamiltonian to the ground state of the final Hamiltonian after a sudden change of the potential.
Similar phenomena where mismatches of energy spectra prohibit transport have been discussed in mass transport \cite{Chien:2013ef} and energy transport \cite{McKay:2013hd}, and later on we will present similar results in the lattice case.

For the dynamically emerged sink at the center, there are less particles flow into the sink when the interaction increases, as shown in Fig.~\ref{fig:GP}f.
On the other hand, more particles can flow into the sink as the interaction strength increases if it is located at the edge.
This subtle difference can be understood from the density distribution of the initial ground state.
As the interaction becomes stronger, the density distribution of the initial ground state without a sink becomes more flat at the center and has relatively more particles towards the edge.
Thus, the density at the sink potential at the edge (center) increases (decreases), as illustrated in Fig.~\ref{fig:GP}c.

The ineffectiveness of a dynamically emerged sink may also be understood from the wave nature of quantum systems.
It is known that when an electromagnetic wave impinges on an aperture whose diameter is much smaller than the wavelength, the transmission is severely suppressed~\cite{jackson1975classical}.
For the atomic analogue of a sink, the particles may be viewed as matter wave whose wavelength is about the size of the whole system.
The matter wave also has very low transmission into a narrow sink potential as shown in our simulations.

Since the GPE is designed for weakly interacting systems, next we will model the same dynamical process of a lattice model allowing us to analyze dynamics in the strongly interacting regime.
The sink potential in the lattice model corresponds to a sudden decrease of the onsite potential on a selected site.
In this approximation, there is only one bound state on the sink site for a noninteracting lattice system, so this is similar to a delta-function potential~\cite{Messiah:1999qm} in the continuum case, $V_{ext}\!=\!-V_l\delta(x)$.
The delta potential only has one bound state regardless of its potential depth $V_l$.
The physics, as one will see, is qualitatively the same as a square sink potential well in the continuum case.

\subsection{Adiabatic Limit}
The general solution of the time dependent Schr\"{o}dinger equation at time $t$ can be expressed as $|\psi(t)\rangle\!=\!\sum_nc_n(t)\psi_n(t)e^{i\theta_n(t)}$, where $\theta_n(t)\!=\!\frac{i}{\hbar}\int_0^{t}E_n(t^\prime)dt^\prime$.
By solving the Schr\"{o}dinger equation,
\begin{equation}
\dot{c}_m(t)\!=\!-c_m\langle\psi_m|\dot{\psi}_m\rangle-\sum_{n\neq m}c_n\frac{\langle\psi_m|\frac{\partial\mathcal{H}}{\partial t}|\psi_n\rangle}{E_n-E_m}e^{i(\theta_n-\theta_m)}.\nonumber
\end{equation}
According to the adiabatic theorem~\cite{Messiah:1999qm}, the system remains in the ground state if $\frac{\partial\mathcal{H}}{\partial t}$ is extremely small when compared to the energy level spacing divided by the natural time unit of the system, which is $\hbar/E_R$ ($\hbar/J$) for the continuum (lattice) model.
In the continuum model, the recoil energy is $E_R=\frac{\pi^2\hbar^2}{2mL_l^2}$ and determines the energy difference between the lowest-energy levels.
The time required to reach the adiabatic limit is limited by the energy difference, so it is proportional to the square of the system size.

\begin{figure}[ht]
	\begin{center}
		\includegraphics[width=0.9\textwidth]{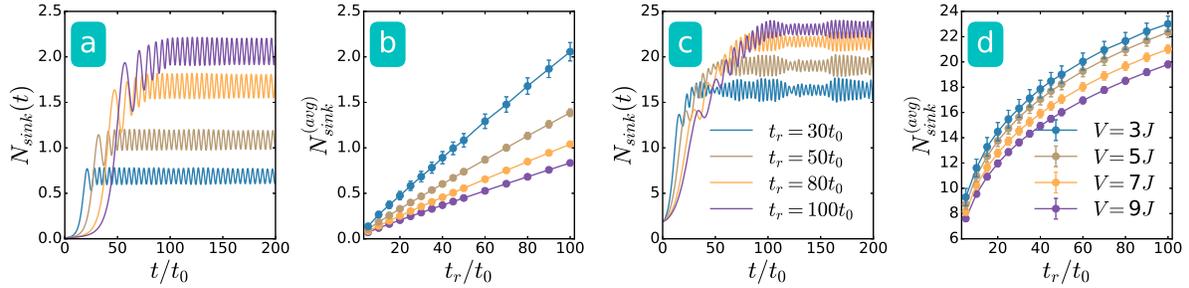}
		\caption{
			Non-interacting bosons in a lattice of $L\! =\! 41$ sites with $N\!=\! 40$ particles. A sink potential appears with ramping time scale $t_r$.
			(a) and (c) show the number of bosons in the sink potential versus time with sink potential depth $V\! =\! 3J$.
			(b) and (d) show the corresponding long-time average versus different ramping times of the sink. The sink is located at the right edge for (a) and (b), and at the center for (c) and (d).
			As $t_r$ increases, the system approaches the adiabatic limit with more particles in the sink.
		}
		\label{fig:Adiabatic1}
	\end{center}
\end{figure}

In Fig.~\ref{fig:Adiabatic1}, we show the dynamics of a small non-interacting lattice system with different ramping times.
For the case with a dynamic sink potential at the center and the case with one at one edge, the sink can accommodate more particles as the ramping time becomes longer.
Moreover, the results show that a dynamic sink with a longer ramping time collects more particles if the sink is at the center ($\sim\!50\%$ in \ref{fig:Adiabatic1}d) than at one edge ($\sim\!5\%$ in \ref{fig:Adiabatic1}b) under the same condition.
We caution that this is again related to the initial density distribution of noninteracting bosons, which is higher at the center and lower at the edge.
In atomtronic applications it is more realistic to consider fast switching of the elements rather than the adiabatic limit, so our main focus is on a sudden emergent (quenched) sink or source potential.

\section{Born-Markov approximations and conserved quantities}
In general, the theoretical framework of open quantum systems consists of a small system (labeled by "s"), which may be the finite lattice considered here, and a large environment (labeled by "e") interacting with the system.
The contribution from the environment is treated as extra terms in the equation of motion of the system.
Such a composite system can be achieved by submerging a lattice system into a background of bosons, and the coupling between them can introduce dissipation or coherent cooling~\cite{Gardiner:2000kq}.
In such a way the system can bypass the conservation of energy.
Recent advances in local heating~\cite{Brantut:2013gs} and single-site cooling~\cite{Grunzweig:2010dr,Kaufman:2012ft} further allow local manipulations to vary the energy of the system.

In order to describe dynamics of open quantum systems, it is more convenient to use the total density-matrix operator $\rho_{\text{total}}$ of the system and the environment.
Tracing over the environment degrees of freedom gives the reduced density matrix of the system.
One usually assumes that initially the system and environment are independent, so $\rho_{\text{total}}\!=\!\rho_{\text{s}}\!\otimes\!\rho_{\text{e}}$ may be used as the initial condition.
In general, the entire open quantum system cannot be solved explicitly due to the large degrees of freedom from the environment.
A manageable description can be obtained with
i) the Born approximation assuming that the frequency scale associated with the coupling between the system and environment is small comparing to the dynamical frequency scales of the system and environment,
ii) the Markov approximation, which requires that the coupling is time-independent over a short time scale and the environment can rapidly return to equilibrium without being altered by the coupling, and
iii) the secular approximation, which discards rapidly oscillating terms in the Markovian master equation.

The Lindblad master equation of an operator $\hat{O}$ in the Heisenberg picture~\cite{breuer2007theory} can be written as
\begin{equation}
\frac{d\hat{O}}{dt}=\frac{i}{\hbar}[\mathcal{H}, \hat{O}]+\gamma\sum_j\left[L_j^\dagger \hat{O} L_j-\frac{1}{2}\{\hat{O}, L_j^\dagger L_j\}\right].
\end{equation}
The observable $\hat{O}$ corresponds to a conserved quantity if it commutes with the Hamiltonian and Lindblad operators.
One can see that by setting $\hat{O}$ to be the total particle number operator, it commutes with the Hamiltonian as well as the particle-number Lindblad operators and the local Lindblad operator we constructed, so the particle number is conserved in those cases.

\end{document}